\documentclass[runningheads]{llncs}
\usepackage[T1]{fontenc}
\usepackage{graphicx}
\usepackage{booktabs}
\usepackage{algorithm}
\usepackage{algpseudocode}
\usepackage[switch]{lineno}

\usepackage{amsmath}

\newtheorem{prob}{Problem Definition}
\begin{document}
\title{Detection of Unknown Errors in Human-Centered Systems}
%
%
\author{Aranyak Maity \and
Ayan Banerjee \and
Sandeep Gupta}
\authorrunning{Maity et al.}

\institute{Arizona State University, USA \\
\email{\{amaity1,abanerj3,sandeep.gupta\}@asu.edu}}
\maketitle              
\begin{abstract}
Artificial Intelligence-enabled systems are increasingly being deployed in real-world safety-critical settings involving human participants. It is vital to ensure the safety of such systems and stop the evolution of the system with error before causing harm to human participants. We propose a model-agnostic approach to detecting unknown errors in such human-centered systems without requiring any knowledge about the error signatures. Our approach employs dynamics-induced hybrid recurrent neural networks (DiH-RNN) for constructing physics-based models from operational data, coupled with conformal inference for assessing errors in the underlying model caused by violations of physical laws, thereby facilitating early detection of unknown errors before unsafe shifts in operational data distribution occur. We evaluate our framework on multiple real-world safety critical systems and show that our technique outperforms the existing state-of-the-art in detecting unknown errors. 
\end{abstract}
\section{Introduction}
\begin{figure}
 \center
\includegraphics[trim=0 0 0 0,width=0.7\columnwidth]{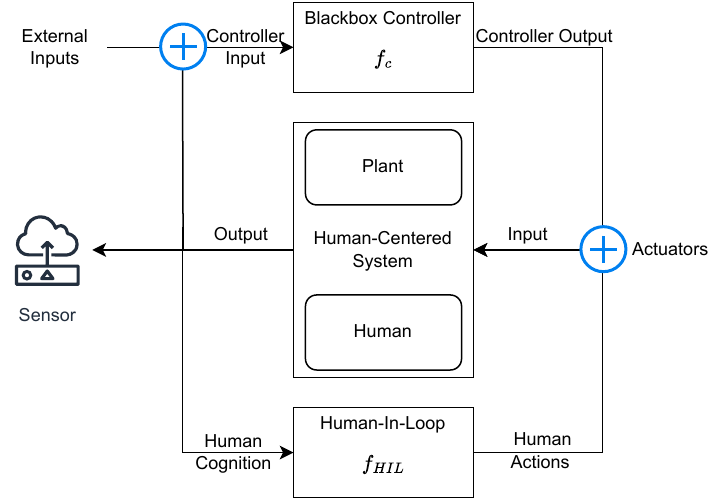} 
 \caption{System Model of Human-Centered Systems. In this architecture, the human operator can be both part of the control mechanism and within the operational dynamics of the plant itself. The plant's state is monitored through sensors and control actions are performed via actuators, processes that are prone to inaccuracies and errors.}
    \label{fig:SystemModel}
\end{figure}

Rapid advancements in Machine Learning and Artificial Intelligence have led to an increase in the number of AI-enabled systems being deployed in real-world safety-critical settings. These systems often are deployed in contexts where they can cause potential risks to human participants. It is of utmost necessity to ensure the safety of such Safety Critical Human-Centered Systems and prevent them from causing harm to humans. While substantial efforts have been made to guarantee the safety of these systems, much of the current research focuses on safety assurances during the design phase~\cite{dreossi2019verifai}\cite{dutta2018learning}\cite{tran2019star}\cite{huang2017safety}\cite{sun2019formal}, often overlooking the unpredictable dynamics of real-world settings and the dynamic nature of the human participants. Additionally, while runtime monitoring has been explored as a solution to this challenge, existing runtime monitoring techniques \cite{weaver2018hybrid} need to be trained on the specific errors they are trying to detect, which are often not available. In this paper,\textbf{we focus on developing an approach for detecting errors in operational Human-Centered Systems, without prior knowledge of the error signatures}. Our approach of error detection relies solely on the observation of inputs and outputs from the system (Figure \ref{fig:SystemModel}). By assuming black-box access (Figure \ref{fig:SystemModel}) to the model's controller, which could be an AI-based controller or a conventional one, such as Model Predictive Control (MPC) we make our detection mechanism model agnostic.

\begin{figure}
\center
\includegraphics[trim=0 0 0 50,width=0.95\columnwidth]{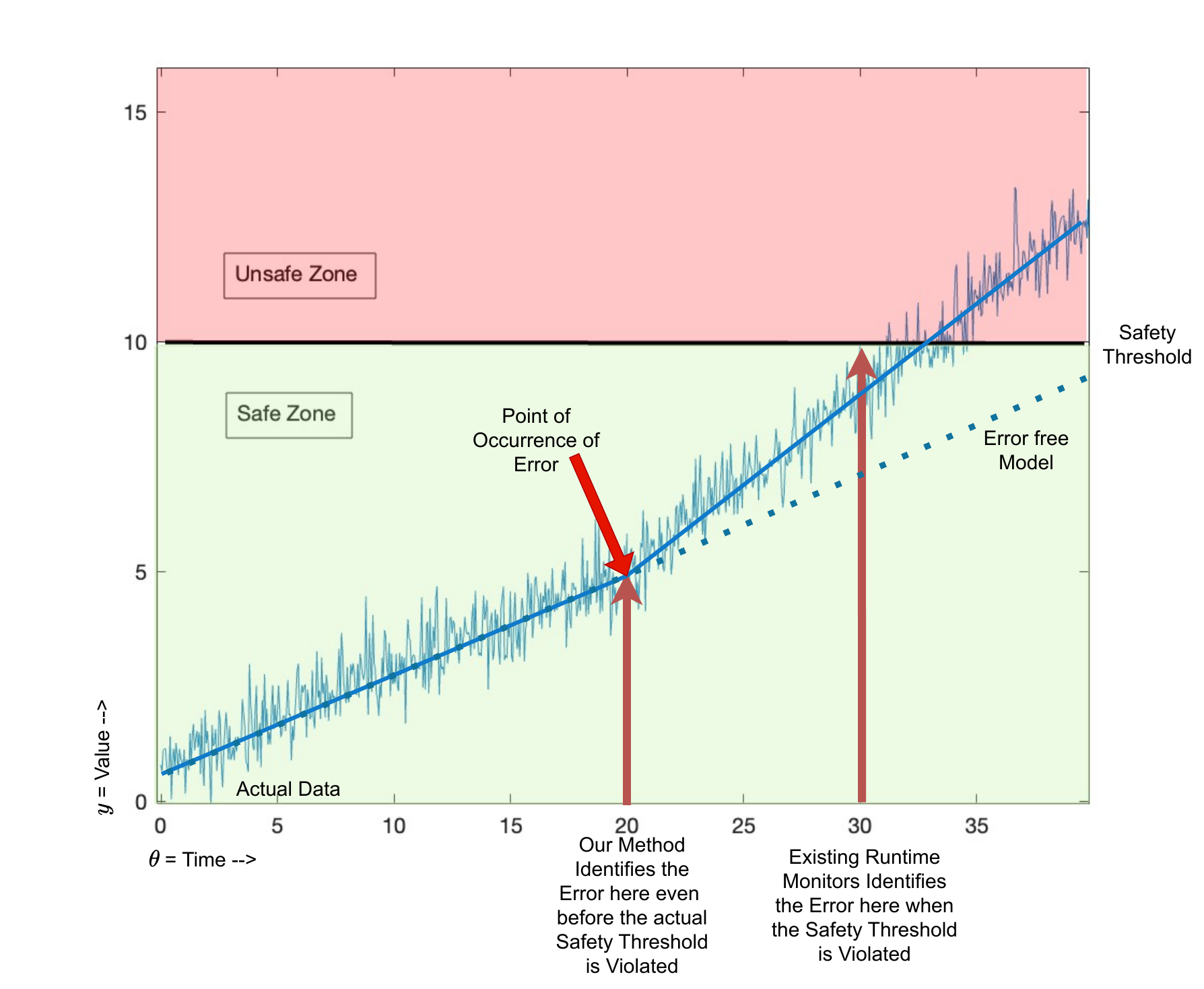} 
 \caption{The figure illustrates a comparative analysis between current runtime monitors and our approach to error detection. While existing techniques can detect errors at 30 when the safety threshold is breached, our approach can identify errors at 20, precisely when they occur. In this example, the input to the system $\theta$ is time and y is the output of the system. }
    \label{fig:toy}
\end{figure}

Recognizing errors in the operational phase presents unique challenges~\cite{banerjee2021faultex}\cite{banerjee2023statistical}\cite{banerjee2023high}. In Human-Centered Systems~\cite{banerjee2024cps} that are in operation, sensing is limited, and also errors in a component of the systems may not readily have any effect on the trajectories of the sensed variables due to several physical properties. Recently proposed design time stochastic safety verification based on output trajectories~\cite{agha2018survey}\cite{kwiatkowska2007stochastic} may fail to detect errors during operation, since the effect of the errors on the output trajectories (sensor values) may fall within the safe operating conditions. An error may subsequently be combined with known or unknown errors resulting in safety violations with potential fatal consequences~\cite{maity2022cyphytest}. Moreover, in real-world deployments, systems may encounter previously unseen scenarios, many of which are unpredictable and lack predefined error signatures, making it challenging to train machines for their detection. Our approach addresses these challenges by deploying continuous model learning and conformance-checking strategy, focused on the model coefficients that reflect the underlying physical laws governing the system. This strategy is designed to identify structural breaks~\cite{safikhani2022joint} and deviations indicative of errors, thereby enhancing error detection without the need for predefined error signatures and contributing to the overall safety of Human-Centered Systems.
 \begin{figure}
\center
\includegraphics[trim=0 0 0 0,width= 0.87\columnwidth]{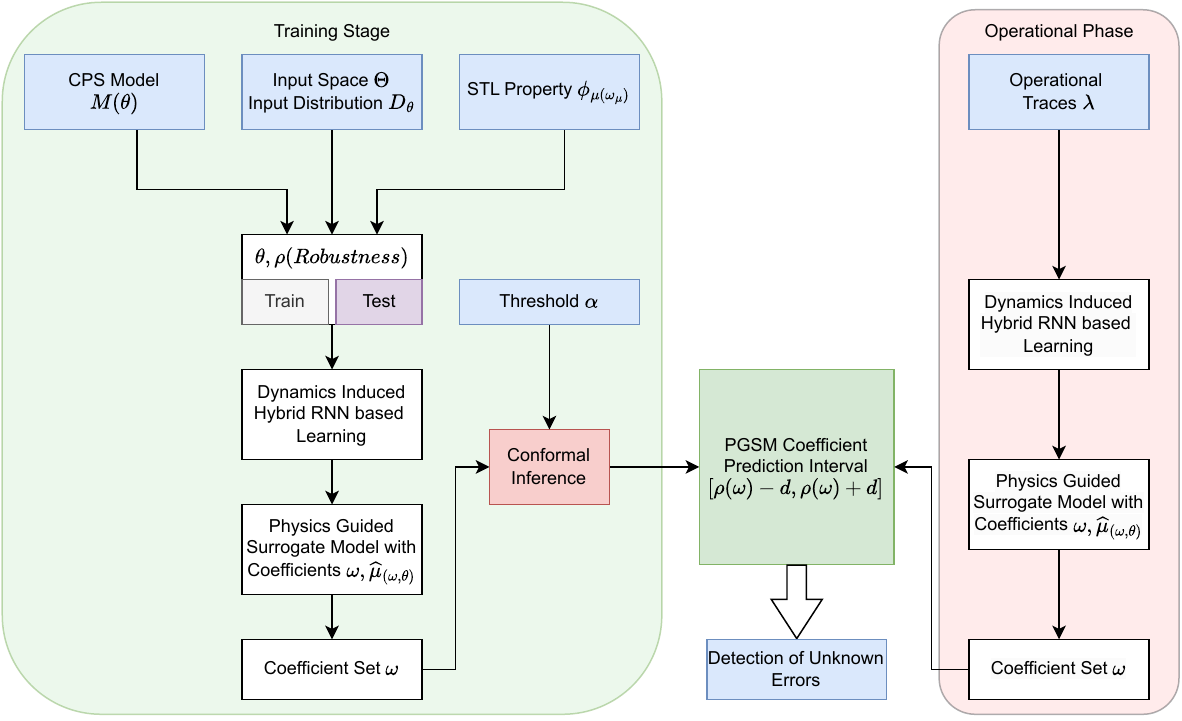} 
 \caption{Overview of the Proposed Approach: 
The diagram illustrates the two-stage process of our methodology. The physics-guided surrogate models facilitate the determination of a conformal range for the surrogate model coefficients. Subsequently, in the Operational Phase, another physics-guided model is learned using real-time operational traces. To ensure the model's conformance, the critical assessment in this phase involves verifying whether the coefficients of this operational model are within the conformal range identified during the training phase.}
    \label{fig:OverviewApp}
\end{figure} 
State-of-the-art error detection uses runtime monitoring and involves learning an operational monitor and testing the conformance of the operational data with the monitor's predictions~\cite{weaver2018hybrid}. An unsafe deviation from the monitor predictions is specified using metric logic such as Signal Temporal Logic (STL)~\cite{qin2022statistical}\cite{qin2024statistical}. The satisfaction of the STL is checked by repeatedly evaluating a robustness value on the operational data~\cite{qin2024statistical}. We illustrate the inadequacy of such an approach in error detection using a toy example shown in Figure \ref{fig:toy} where there is an unknown error at 20. 
State-of-art runtime monitoring technique using conformal inference on operational data~\cite{qin2024statistical}, when implemented in the above example (Figure \ref{fig:toy}), is not able to detect the error when it occurs but detects it at a far later point (at 30.1) when the error has already precipitated into a safety violation. In contrast, implementing our strategy as detailed in this paper, annotates the input segment starting at 20 as unsafe. In our approach, we combine continuous model learning and conformal inference on model coefficients to partition the input space into safe and unsafe regions based on whether the learned model is violating the safety STL on model coefficients. Note that other runtime monitoring or error detection techniques that require the predefined error signatures where even unable to detect the presence of the error as the error is assumed to be an unknown-unknown error~\cite{maity2023detection} and such error signatures are not available. 

Our solution, and core contribution, is the introduction of continuous model learning and conformal inference on model coefficient. Model coefficients represent the relationship between the input and the output trajectories of the system guided by the physics laws. If an unknown-unknown error affects the system it will lead to inaccurate or deviating model coefficients. This is because the model encapsulates the relationship between the input and the output trajectories and if there is an error it would lead to different model coefficients to compensate for the changes in the system. So in this paper, we propose a model conference on model coefficients rather than on the output trajectories. We show that by converting the STL on model coefficients it is able to detect unknown errors in Human-Centered systems without the need for predefined error signatures.

Our approach is a two-stage process (see Figure \ref{fig:OverviewApp}), 1) In \textbf{Training Stage} - we learn physics-guided surrogate models to determine a conformal range for safe operation on the model coefficients and 2) In \textbf{Operational Phase} -  we relearn the physics guided surrogate model and check conformance of the model coefficients to determine the existence of errors in the operational traces. Through a series of real-world error detection experiments, we show that a) our method can detect errors even when error signatures are unavailable, b)the technique is model agnostic that is it doesn't depend on the specific system model of the human-centered system and, c)enables early detection of unknown errors whereby enabling safe operations of these systems.
\subsection{Contributions}
In this paper, we make the following contributions:
\begin{itemize}
\item Provide a generic framework for stochastic model conformance checking on model coefficients and not on output trajectories.
\item Provide a mechanism to mine physics-guided operational models from operational traces of Human-Centered Black Box Systems.
\item Show detection of errors in the artificial pancreas, autonomous vehicles, and aircraft examples.
\end{itemize}

\subsection{Paper Organization}
The paper is organized in the following pattern. Section 2 defines the required preliminaries and background work. Section 3 explains the methodology for mining the model coefficients. Section 4 explains how model conformance can be utilized on the model coefficients derived from Section 3. Section 5 discusses the case studies we use to verify the proposed method.  Section 6 explains the evaluation criteria and Section 7 shows the results of the analysis performed on the examples defined in Section 5.

\section{Preliminaries}
\textbf{Physics Model:} A physics model is a dynamical system expressed using a system of linear time-invariant ordinary differential equations in Eqn. \ref{eqn:set}. The system has $n$ variables $x_i$, $i \in \{1\ldots n\}$ in an $n \times 1$ vector $\mathcal{X}$, $\mathcal{A}$ is an $n \times n$ coefficient matrix, $\mathcal{B}$ is an $n \times n$ diagonal coefficient matrix.
\begin{equation}
\label{eqn:set}
\dot{X(t)} = \mathcal{A}X(t) + \mathcal{B}U(t),\text{  } Y(t) = \beta X(t) 
\end{equation}
where $U(t)$ is a $n \times 1$ vector of external inputs. $Y(t)$ is the $n \times 1$ output vector of the system of equations. An $n \times n$ diagonal matrix, $\mathcal{\beta}$ of 1s and 0s, where $\beta_{ii} = 1$ indicates that the variable $x_i$ is an observable output else it is hidden and is not available for sensing.

A formal object $\hat{\mu}$ is a physics model when the set of models $\mu$ can be described using the coefficient $\omega =\mathcal{A} \bigcup \mathcal{B}$. The formal object can then take any $\theta$ as input and given the model coefficients $\omega$, generate a trace $\zeta_\theta = \hat{\mu}(\omega,\theta)$. 

\textbf{Trajectory and Models:} 
A trajectory $\zeta$ is a function from a set $[0, T]$ for some $T \in \mathcal{R}^{\geq 0}$ denoting time to a compact set of values $\in \mathcal{R}$. The value of a trajectory at time $t$ is denoted as $\zeta(t)$. Each trajectory is the output of a model $M$. A model $M$ is a function that maps a $k$ dimensional input $\theta$ from the input space $\Theta \subset \mathcal{R}^k$ to an output trajectory $\zeta_\theta$.     

The input $\theta \in \Theta$ is a random variable that follows a distribution $\mathcal{D}_\Theta$. The model $M$, can be simulated for input $\theta$ and a finite sequence of time $t_0 \ldots t_n$ with $n$ time steps and generate the trajectory $\zeta_\theta$ such that $\zeta_\theta(t_i) = \Sigma(\theta,t_i)$.

\textbf{Trace:}
Concatenation of $p$ output trajectories over time $\zeta_{\theta_1} \zeta_{\theta_2} \ldots \zeta_{\theta_p}$ is a trace $\mathcal{T}$.

\textbf{Continuous model mining:} 
Given a trace $\mathcal{T}$, continuous model mining maps the trace into a sequence $\Omega$ of $p$, $\omega_i$s such that $\forall i$  $dist(\hat{\mu}(\omega_i,\theta_i),\zeta_{\theta_i}) < \upsilon$, where $dist(.)$ is a distance metric between trajectories and $\upsilon \approx 0$ is decided by the user.

 \subsection{Signal Temporal Logic}
  Signal temporal logic are formulas defined over trace $\mathcal{T}$ of the form $f(\Omega) \geq c$ or $f(\Omega) \leq c$. Here $f: \mathcal{R}^p \rightarrow \mathcal{R}$ is a real-valued function and $c \in \mathcal{R}$. STL supports operations as shown in Eqn. \ref{eqn:STL}.
\begin{equation}
\label{eqn:STL}
\phi, \psi := true | f(\Omega) \geq c | f(\Omega) \leq c | \neg \phi | \phi \wedge \psi | \phi \vee \psi|F_I \phi | G_I \phi | \phi U_I \psi,
\end{equation}  
   where $I$ is a time interval, and $F_I$, $G_I$, and $U_I$ are eventually, globally, and until operations and are used according to the standard definitions\cite{donze2010robust}\cite{fainekos2009robustness}.
To compute a degree of satisfaction of the STL we consider the robustness metric.
   The \textbf{robustness value} $\rho$ maps an STL $\phi$, the trajectory $\zeta$, and a time $t \in [0, T]$ to a real value. An example robustness $\rho$ for the STL $\phi: f(\Omega) \geq c$ is $\rho(f(\Omega)\geq c,\Omega,t) = f(\Omega(t)) - c$.
       
\subsection{Physics-Driven Surrogate Model }
A surrogate model is a quantitative abstraction of the black box model $M$. A quantitative abstraction satisfies a given property on the output trajectory of the model. In this paper, this quantitative property is the robustness value of an STL property. With this setting, we define a $\delta$-surrogate model $\hat{\mu}$.  \\
\textbf{($\delta$, $\epsilon$) Probabilistic Surrogate model:} Let $\zeta_\theta$ be a trajectory obtained by simulating $M$ with input $\theta$. Let $\omega^T$ be the coefficients of the physics-guided representation of the original model. Given a user-specified $\epsilon$, the formal object $\hat{\mu}(\omega,\theta)$ is a ($\delta$,$\epsilon$) probabilistic distance preserving surrogate model if 
\begin{equation}
\label{def:DE} 
\exists \delta \in \mathcal{R}, \epsilon \in [0,1] : P(|\rho(\phi,\omega^T)-\rho(\phi,\omega)| \leq \delta) \geq 1-\epsilon.
\end{equation}
A $\delta$ surrogate model guarantees that the robustness value evaluated on a physics model coefficient $\omega$ derived from the trajectory $\zeta_\theta$ will not be more than $\delta$ away from the robustness computed on the coefficients of the original model $M$.

\section{Coefficient Mining from Trajectory}
\begin{prob}
\label{prob:Prob}
Given a set of variables $\mathcal{X}(t)$, a set of inputs $U(t)$, a $\beta$ vector indicating observability, and a set $\mathcal{T}$ of traces such that $\forall i : \beta_i = 1 \exists T(x_i) \in \mathcal{T} $ and $\forall u_j(t) \in U(t) \exists T(u_j) \in \mathcal{T}$.
\begin{algorithm}
	\caption{RNN induction algorithm}
\begin{algorithmic}[1]
		\State $\forall x_i \in X$ create an RNN node with $n + 1$ inputs and  $x_i$ as the hidden output.   
		\For {each RNN node corresponding to $x_i$}
		 	\For {each $j \in {1 \ldots n}$}  
				\If {$a_{ij} \neq 0$}
					\State Add a connection from the output of RNN node for $x_j$ to the input of RNN node for $x_i$.				
				\EndIf
			\EndFor	
			\State Remove all other inputs in the RNN which does not have any connection.
			
			\For each $j \in {1 \ldots n}$  
				\If {$b_{ij} \neq 0$}
					\State Add $u_j$ as an external input to the RNN node for $x_i$.				
				\EndIf
			\EndFor				
			
		\EndFor
		\State Assign arbitrary weights to each link.
\end{algorithmic}
	\label{alg:Induction}
\end{algorithm}
\noindent{\bf Derive:} approximate coefficients $\mathcal{A}^a$ and $\mathcal{B}^a$ such that: 
\begin{itemize}
\item $\forall i,j$ $|\mathcal{A}^a(i,j)-\mathcal{A}(i,j)| < \xi$
\item $\forall i$ $|\mathcal{B}^a(i,i)-\mathcal{B}(i,i)| < \xi$
\item Let $\mathcal{T}^a$ be the set of traces that include variables derived from the solution to differential equation $\frac{dX(t)}{dt} = \mathcal{A}^aX(t) + \mathcal{B}^aU(t)$ then $\forall i : \theta_i = 1,$ and $\forall k \in \{1 \ldots N\}, |T^a(x_i)[k] - T(x_i)[k]| < \Psi T(x_i)[k]$,
\end{itemize}   
where $\xi$ is the error in the coefficient estimator, while $\Psi$ is the error factor for replicating the traces of variables with the estimated coefficients. 
\end{prob}

\subsection{Dynamics Induced RNN}
\label{sec:dihrnn}
\begin{figure}
\center
\includegraphics[trim=0 0 0 0,width=\columnwidth]{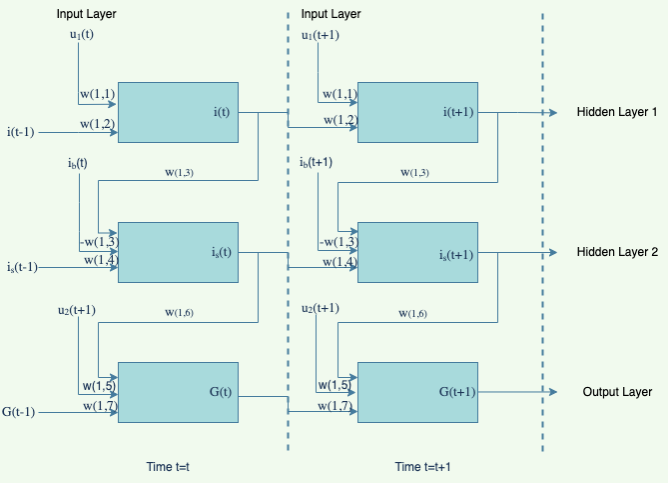} 
 \caption{DiHRNN structure of the Bergman Minimal Model}
    \label{fig:Implementation}
\end{figure}

For each variable $x_i \in X$ the system of dynamical equations takes the form in Equation \ref{eqn:ind}.
\begin{equation}
\label{eqn:ind}
\frac{dx_i}{dt} = \sum_{j = 1}^{n}{a_{ij} x_j} + b_{ii}u_i.
\end{equation}
The RNN induced by the system of equations (Equation \ref{eqn:set}) follows Algorithm \ref{alg:Induction}.
We explain Algorithm \ref{alg:Induction} using the linearized Bergman Minimal Model (BMM) as an example. The model is a dynamical system that mimics the glucose-insulin biochemical dynamics in the human body. The Bergman Minimal model is linearized using Taylor Series expansion starting from overnight glucose dynamics and going up to time $N$. The linearized model is represented in Equation \ref{eqn:8}, \ref{eqn:9} and \ref{eqn:10}.
\begin{eqnarray}
\label{eqn:8}
& &\frac{d\delta i(t)}{dt} = -n\delta i(t) + p_4 u_1(t)\\
\label{eqn:9}
& & \frac{d\delta i_s(t)}{dt} = -p_1 \delta i_s(t) + p_2 (\delta i(t) - i_b)\\
\label{eqn:10}
& & \frac{d\delta G(t)}{dt} = -\delta i_s (t) G_b -p3 (\delta G(t)) + u2(t)/VoI,
\end{eqnarray}
The input vector $U(t)$ consists of the overnight basal insulin level $i_1b$ and the glucose appearance rate in the body $u_2$. The output vector $Y(t)$ comprises the blood insulin level $i$, the interstitial insulin level $i_s$, and the blood glucose level $G$. For this example, we consider that only the blood glucose level $G$ is an observable output of the system of equations. $i_s$ and $i$ are intermediate outputs that are not measurable for the system of equations and only contribute to the final glucose output. $p_1$, $p_2$, $p_3$, $p_4$, $n$, and $1/V_o I$ are all the coefficients of the set of differential equations. The resulting DiH-RNN for the BMM using Algorithm \ref{alg:Induction} is shown in figure \ref{fig:Implementation}

\subsection{Forward pass in DiH-RNN}
We prove that the Forward pass on an RNN node in Equation \ref{eqn:hidden} estimates the solution of Equation \ref{eqn:ind} with error factor $\Psi$ if $\tau \leq min_{i}{\frac{\sqrt{2\Psi}}{a_{ii}}}$.
The poof is attached in the appendix.

\subsection{Backpropagation to learn coefficients}
The main aim of backpropagation is to derive the approximate coefficient matrices $\mathcal{A}^a$ and $\mathcal{B}^a$. Given an error ratio of $\phi$, we have established that the forward pass is convergent and estimation error is proportional to $\phi$ if $\tau \leq \frac{\sqrt{2\Psi}}{|a_{ii}|} \forall i$. However, we do not know $a_{ii}$ and hence setting $\tau$ is a difficult task. Often $\tau$ is limited by the sampling frequency of the sensor. In this paper, we assume that the $\tau$ satisfies the condition for convergence of the forward pass. Proposition 1 in \cite{yuan2021physics} shows that for shallow DNNs if all the weights are nonnegative and the activation function is convex and non-decreasing then the overall loss is convex. In such a scenario there exists only a single minima and the gradient descent mechanism is guaranteed to find it.

\section{Conformal Inference}
 
Conformal Inference is a framework to predict the accuracy of the predictions in a regression framework Conformal Inference is rigorously studied in the following works~\cite{maity2023detection,qin2022statistical,qin2024statistical}. We use that basic framework of conformal inference and extend it to model coefficients. 
In this approach (Figure \ref{fig:OverviewApp}), we use error-free operational traces to learn the confidence threshold d. The process to determine this threshold on model coefficients involves several steps:\\
1) Split the error-free data into training and testing sets. \\
2) From the training set, calculate a set of PGSM model coefficients, $\omega_e$.\\
3) For each subset in the testing set, compute model coefficients, $\omega_i$, where i represents the specific subset. \\
4) Using the $\omega_e$ from the train set we calculate the robustness residue of each test $\omega_i$. We define robustness as a quantification of the difference in model coefficient values. For this paper, we consider the maximum deviation of the model coefficients which is explained by the following equation \ref{eqn:phi}. Other metrics like minimum deviations and average deviations could have been used but such investigations are beyond the scope of this paper and are left for future investigation.
\begin{equation}
\label{eqn:phi}
\rho(\phi,\omega) = max_{j \in \{1 \ldots n\}}{abs({(\omega[j] - \omega_e[j])}/{\omega_{e}[j]})} - \alpha, 
\end{equation}
where n is the total number of model coefficients in $\omega_e$.\\
5) Sort the calculated robustness values in ascending order and identify the residue corresponding to the position defined by $\lceil(n/2 +1)(1-\alpha)\rceil$.\\
6) The robustness residual value at the given position gives us the confidence range d, and with it, we derive the confidence interval $[\rho(\omega)-d,\rho(\omega)+d]$.

Any new data with unknown errors should result in model coefficients such that the STL robustness residue is beyond the range $[\rho(\omega)-d,\rho(\omega)+d]$.


\section{Case Studies}
Human-centered systems are those where failure could result in catastrophic outcomes, such as loss of life, significant property damage, or harm to the human participant. In this section, we present three real-world safety critical examples. Each example features a human integrated into the operational dynamics, as outlined by the architecture depicted in Fig \ref{fig:SystemModel}. The inclusion of humans within the operational framework elevates the criticality of these systems, significantly increasing the risk of harm. In these cases, the problem of detection of unknown errors is even more important.

\subsection{Automated Insulin Delivery System Example}
\label{sec:Example}
\begin{table*}[t]
	\centering
	\scriptsize
	\caption{Physical model coefficients derived using DiHRNN for train and test set}
	\begin{tabular}{p{1.0 in}|p{0.4 in}|p{0.4 in}|p{0.6 in}|p{0.3 in}|p{0.5 in}|p{0.4 in}|p{0.48 in}|p{0.35 in}}
	 \toprule
		{Train / Test} & {$p_1$ 1/min} & { $p_2$ 1/min} &{$p_3 $ $\frac{10^{-6}}{\mu U.min^2}$} & $p_4$ & {$n$ 1/min} & {$VoI$ dl} & $G_b$ mg/dl & Residue\\ \midrule
Simulation Settings & 0.098 &0.035&	0.028 &	0.05 & 0.1406 &199.6 & -80 & NA \\
Train & 0.0978 &0.0349 & 0.0262 &	0.0508 & 0.1406 & 198.134& 	-80.64 & 0\\
Test 1 & 0.0982   & 0.0329 &  0.0256   & 0.0530 & 0.1405  &198.1340&  -80.2774 & 0.0225\\
2 &     0.0979  &0.0332 &   0.0274 &   0.0533 & 0.1407 &198.1340 &  -85.0589 & 0.0028\\
3 &     0.0980 & 0.0348 &   0.0262 &   0.0528 &  0.1405 &198.1340 & -85.0973&0.0011\\
4 &     0.0981  &0.0343 &   0.0267 &   0.0515 &  0.1405 &198.1340 &   -80.6921 & -0.0168 \\
5 &    0.0979  & 0.0317&   0.0273 &   0.0548 & 0.1407 &198.1340  &  -82.7676& 0.0328 \\
6 &     0.0980  &0.0328  &   0.0275 &   0.0534 &  0.1404 &198.1340 &   -82.3447& 0.0048\\

		\bottomrule
	\end{tabular}
	\label{tbl:Ex2}
	\vspace{-0.1 in}
\end{table*}
In the Automated Insulin Delivery (AID) system, the glucose-insulin dynamics is given by the Bergman Minimal Model (BMM) represented as \ref{eqn:8}, \ref{eqn:9}, and \ref{eqn:10} and is explained in detail in \ref{sec:dihrnn}.
For this paper, we consider the unknown error of insulin cartridge error in the automated insulin delivery system. The error signature of the error was unavailable at the time of the error as this error was never seen before. The human being part of the system being controlled made measuring the effects even more complicated.  While the controller operated under the assumption of flawless insulin administration, the actual delivery to the human body (the system) was compromised, leading to a significant disparity between the system's state as perceived by the controller and its true state.

\subsection{Aircraft Example}
\label{sec:ExampleAOA}
Pitch control in an aircraft is automated using a Proportional Integrative Derivative (PID) Controller. The pitch control system considers a linear system model described by Equation \ref{eqn:Aero} \cite{messner2020control}.
\begin{eqnarray}
\label{eqn:Aero}
\dot{x_\alpha} &=& c_{\alpha \alpha}  x_\alpha + c_{\alpha q}  x_q + c_{\alpha \delta}  u_\delta \textnormal{,   } \dot{x_q} = c_{q \alpha}  x_\alpha - c_{qq}  x_q + c_{q \delta}  u_\delta\\\nonumber
\dot{x_\theta} &=& c_{\theta \theta} x_q, \textnormal{ } y(t) = x_\theta.
\end{eqnarray}

Here $x_\alpha$ is the angle of attack (AoA), $x_q$ is the pitch rate, $u_\delta$ is the elevator angle, and $x_\theta$ is the pitch angle of the aircraft. The controller is a PID and based on a pitch angle set point derives the elevator angle $u_\delta$. Hence, $u_\delta$ is the input to the aircraft dynamics, while $x_\theta$ is the output of the dynamical model. A trajectory is the continuous time value of state variables in between two elevator angle inputs from the PID.

For this example, we consider the unknown MCAS error that caused the accidents in the Boeing 787 aircraft. The cause was also unknown due to the black box abstraction of the MCAS system. The human participant (the pilot in this case) did not know that the faulty AOA sensor was being used to control the plant. 

\subsection{Autonomous Driving Example}
\label{sec:ExampleAutoVehi}
An autonomous car detects another static car in its lane and attempts to stop before crashing into the car ahead. The kinematics of the car is given by:
\begin{eqnarray}
    & \dot{a_x} = -0.01s_x+0.737-0.3v_x-0.5a_x \textnormal{,} \\\nonumber
    &\dot{v_x} = 0.1 a_x \textnormal{,  } \dot{s_x} = v_x - 2.5 \textnormal{.}
\end{eqnarray}  
For this example, we consider the unknown error of a zero-day vulnerability in the controller code. The vulnerability caused the black box controller code to change from $f_c$ to ${f^\prime}_c$ in Fig \ref{fig:SystemModel}.  Originating from a zero-day vulnerability, the full impact on the system was uncertain, given that this vulnerability had not been detected before.

\section{Evaluation Method and Metrics}
Human-centered systems are safety-critical, and it is necessary to identify unknown errors to shield the human participant from harm. The performance of zero-shot detection of unknown errors is quantified in terms of the true positive rate of the detection algorithm. We designate the approach as Detected (D) if it can identify the Unknown-Unknowns, and Undetected (ND) if it cannot. The availability of real data for such real-world safety-critical systems with unknown errors is fairly limited. So, here we use simulators developed in MATLAB to generate data for such unknown errors in real-world complex systems.

\subsection{Unknown-Unknown Scenario Simulation}
For the AID example, we use the shunted insulin model to generate the traces with the insulin cartridge errors. We vary the amount of insulin blockade percent between 20 to 80 percent and the time until insulin release from 50 to 150 mins. The scenarios generated for the insulin cartridge problem are listed in Table \ref{tbl:humanEx}. 

For the AoA error in the MCAS system, we use any error or noise rate of 20-25\% in the AoA measurement and use that to derive the coefficients at the model of the pitch control system.

For the autonomous vehicle example, an integer overflow vulnerability in the control software is considered where instead of declaring Q as an \texttt{uint16\_t} variable it is mistakenly defined as \texttt{int8\_t}. This means that instead of setting $Q(1,1) = 10,000$, it is now set at $Q(1,1) = 16$. This can potentially cause a crash since the controller is less aggressive.

\subsection{Baseline Strategy}
We replicate the model conformance-based strategy described in \cite{qin2022statistical}\cite{qin2024statistical} to the best of our knowledge. In the work, the authors learn a surrogate model of the system under test and use it to find the robustness range of the output values. During operation, a new model is learned from the test traces and checked if the robustness values lie within the robustness range. If the robustness value of the test system is outside the range then the system under test is termed to have deviated from the approved characteristics.
\section{Results}
In safety-critical systems that involve human participants, it is of utmost necessity to detect every possible error to stop the faulty system from causing any harm to humans. It is established that detecting such errors that is the number of true positives detected is far more important than other metrics. So, in this paper, we consider the true positive rate of the different detection algorithms.
\subsection{Automated Insulin Delivery System Example}
Table \ref{tbl:humanEx} shows that for the insulin cartridge problem, the model conformance results show that the robustness values under various input configurations are falling outside the range. Hence, these scenarios are deemed to be non-conformal to the original model. Using the technique defined in this paper, it was able to detect all the unknown errors simulated for evaluation purposes without the need for error signatures and have a positive predictive value of 100\%.

\begin{table*}[t]
	\centering
        \scriptsize
	\caption{Comparison of physical model coefficients derived using DiH-RNN for different Insulin Blockages to detect the errors, D in the robustness column means error detected and Robustness is beyond [-0.0216, 0.0376]. Insulin = 7.5 U, Meal = 20 grams. }
	\begin{tabular}{p{0.6 in}|p{0.3 in }|p{0.3 in}|p{0.3 in}|p{0.4 in}|p{0.3 in}|p{0.3 in}|p{0.35 in}|p{0.35 in}|p{0.3 in}|p{0.49 in}|p{0.4 in}}
	 \toprule
		{Insulin Block Percentage} &{Time until insulin release}& {$p_1$ 1/min} & { $p_2$ 1/min} &{$p_3 $ $\frac{10^{-6}}{\mu U.min^2}$} & $p_4$ & {$n$ 1/min} & {$VoI$ dl} & $G_b$ mg/dl & Robu-stness & Model Conformance on coefficients(Our Method) & Model Conformance on Output \cite{qin2024statistical,qin2022statistical}\\ \midrule
20 & 150 &0.098 &0.033& 0.018 &0.065  &0.1404 &268.55&-51.46& 0.37 & (D) & ND\\
40 & 120 & 0.098&0.034& 0.018 &0.053 &0.1402& 287.92 &  -68.32& 0.3885 &(D) &ND\\
80 & 90 & 0.098 &0.034&0.019&0.068 &0.1401 &235.25& -58.68 & 0.36 &(D)&ND\\
70 & 70 & 0.098&0.033& 0.020 &  0.068 &0.1400&216.14& -48.12& 0.43 &(D) & ND\\
60 & 50 & 0.098&0.034& 0.019 &0.068 &0.1405&180.48 &-69.76& 0.35 &(D)& ND \\
Phantom 20 & 150 & 0.098 & 0.0269& 0.0194 & 0.058 &0.1402&155.89& -54.104 & 0.32 &(D)&ND\\
Phantom 40 & 120 & 0.098&0.0339&	0.0218&	0.0579&0.1402&307.06&	-60.73 & 0.5284 &(D)&ND\\
Phantom 80 & 90 & 0.098&0.0344& 0.0217&	0.0503&0.1401&143.43&-64 & 0.27 &(D)&ND\\
Phantom 70 & 70 & 0.098&0.0348&	0.0229&	0.0655&0.139&	169.20	&-48.26&0.48 &(D)&ND\\
Phantom 60 & 50 & 0.0983&0.0349&	0.0187&	0.0554&0.1400&317.86& -55.12 & 0.5825 &(D)&ND\\
		\bottomrule
	\end{tabular}
	\label{tbl:humanEx}
	\vspace{-0.15 in}
\end{table*}
\subsection{Aircraft Example}
As shown in Table \ref{tbl:aoaExV2}, the model conformance with STL on the model outputs failed to recognize errors as the outputs fell within the defined safe and robust range. In contrast, our proposed detection technique by applying model conformance to the model's parameters, successfully identified 9 out of 10 such errors immediately upon their occurrence and had a positive predictive value of 90\%.
\begin{table*}[t]
	\scriptsize
	\centering
	\caption{Comparison of physical model coefficients derived using DiH-RNN for different AoA errors and error timings to detect errors, D in the robustness column means error detected and Robustness is beyond [0.0299, 0.1116].}
	\begin{tabular}{p{0.3 in}|p{0.32 in }|p{0.22 in}|p{0.25 in}|p{0.3 in}|p{0.2 in}|p{0.25 in}|p{0.35 in}|p{0.3 in}|p{0.3 in}|p{0.25 in}|p{0.39 in}|p{0.48 in}|p{0.37 in}}
	 \toprule
		{Set point - SP (rads)} &{SP change time (s)}& AoA error (rad)& Error Time (s) & {$c_{\alpha \alpha}$ 1/sec} & { $c_{\alpha q}$} &{$c_{\alpha \delta}$ 1/sec} & {$c_{q \alpha}$ 1/sec} & {$c_{q q}$ 1/sec} & {$c_{q \delta}$ 1/$sec^2$} & $c_{\theta q}$ 1/sec & Robust-ness& Model Conformance on coefficients (Our Method) & Model Conformance on Output \cite{qin2024statistical,qin2022statistical}\\ \midrule
0.2 & 0 & 0.6 & 5 & -0.276 & 53.7 & 0.24 & -0.0118 & -0.475 & 0.0232 & 60.1 & 0.136 &(D)&ND\\
0.5 & 5 & 0.2 & 7 & -0.258 & 47.6 & 0.24 & -0.0123 & -0.482 & 0.0205 & 62.9 & 0.156 &(D)&ND\\
0.4 & 2 & 0.4 & 10 & -0.282 &45.5& 0.24 & -0.0115 & -0.51 & 0.0213 & 66.11 & 0.22 &(D)&ND\\
0.8 & 5 & 0.4 & 5 & -0.281 & 60.2 & 0.27 & -0.0126 & -0.4 & 0.0219 & 65.2 & 0.13 &(D)&ND\\
0.1 & 5 & 0.6 & 5 & -0.269 & 52.4 & 0.25 & -0.0129 & -0.489 & 0.0231 & 63.2 & 0.17 &(D)&ND \\
0.1 & 7 & 0.6 & 5 & -0.28 & 63.6 & 0.26 & -0.0123 &	-0.367 & 0.0214 & 60.23 & 0.11 &(ND)&ND \\
0.1 & 9 & 0.7 & 2 & -0.257 & 65.0 &	0.27 &	-0.0136 &	-0.354&	0.0219 &	58.3 & 0.16 &(D)&ND\\
0.4 & 9 & 0.9 & 2 & -0.26&	64.8& 0.25&	-0.0138&	-0.398 &	0.0243&	61.2 & 0.17 &(D)&ND\\
0.1 & 10 & 0.6 & 10 & -0.293 &67.5&	0.27&	-0.0136 &	-0.372 &	0.023 &	66 & 0.17 &(D)&ND\\
0.3 & 1 & 0.3 & 10 & -0.302 & 46.1 &	0.28 &	-0.0125 &	-0.46 &	0.0214 & 63.3 & 0.18 &(D)&ND\\
		\bottomrule
	\end{tabular}
	\label{tbl:aoaExV2}
	\vspace{-0.2 in}
\end{table*}
\subsection{Autonomous Driving Example}
We conducted 11 simulations of the autonomous braking system, using the data to train a deep learning model for assessing the reliability of the system's output. Subsequently, we carried out an additional 11 simulations introducing braking errors. The vulnerable controller code was executed to obtain the traces starting from the same initial $s_x$ and $v_x$ as training. The average robustness residue is -17.395 ($\pm$ 2.1 ), with all vulnerable traces falling outside the robustness range. Our proposed method using DiH-RNN, implementing STL on the model's parameters, detected all 11 errors and had a positive predictive value of 100\%.
\section{Conclusions}
This paper proposes a model-agnostic framework for the detection of unknown errors in operational human-centered systems without the need for error signatures. By employing a physics-guided surrogate model to track the physical system's behavior and using a hybrid RNN approach to derive model coefficients, our method identifies deviations using conformal inference techniques, signaling unknown errors in the operational system. With our technique, we can detect errors that haven't been identified before and can stop the system from causing harm to the human participant. Our results demonstrate that this method surpasses existing state-of-the-art error detection techniques in identifying errors without relying on pre-established error definitions. However, it's important to note that our method's efficacy relies heavily on the training and testing data, particularly when determining the model coefficients. Furthermore, the impact of available training data on error detection accuracy needs further investigation.

%
%
%
\bibliographystyle{splncs04}
\bibliography{demo,demo1}

\end{document}